\documentclass[twocolumn,pre,showpacs,showkeys,nofootinbib]{revtex4}
\usepackage{graphics,graphicx}

\begin{document}
\title{Weighted Assortative And Disassortative Networks Model}
\author{C.~C. Leung}
\author{H.~F. Chau\footnote{Corresponding author, electronic address:
 {\tt hfchau@hkusua.hku.hk}}}
\affiliation{Department of Physics, University of Hong Kong, Pokfulam Road,
 Hong Kong}
\affiliation{Center of Theoretical and Computational Physics, University of
 Hong Kong, Pokfulam Road, Hong Kong}
\date{\today}

\begin{abstract}
Real-world networks process structured connections since they have non-trivial
vertex degree correlation and clustering.
Here we propose a toy model of structure formation in
real-world weighted network. In our model, a network evolves by topological
growth as well as by weight change.
In addition, we introduce the weighted assortativity coefficient,
which generalizes the assortativity coefficient of a topological network,
to measure the
tendency of having a high-weighted link between two vertices of similar
degrees.
Network generated by our model exhibits scale-free behavior with a tunable
exponent. Besides, a few non-trivial features found in real-world networks
are reproduced by varying the parameter ruling the speed of weight evolution.
Most importantly, by studying the weighted assortativity coefficient,
we found that both topologically assortative and disassortative networks
generated by our model are in fact weighted assortative.
\end{abstract}

\maketitle
\section{Introduction} \label{intro}
There is a wave of interest to study complex networks by 
statistical physical means \cite{review1,review2,review3}. The area of
studies include the Internet \cite{internet, internet2, internet3}, the
World-Wide Web \cite{www}, 
scientific collaboration networks (SCN) \cite{scn,scn2}, biological networks 
\cite{bio,bio2} and the world-wide airport network (WAN) 
\cite{airport}.
In all these cases, one can naturally identify the subject under
study with a graph.
For instance, an airport in a WAN can be represented by a vertex; and there is
a link between two 
vertices if and only if there is a direct flight between the corresponding 
airports. Although these complex networks are drawn from 
vastly different systems in the real
world, they exhibit the following universal properties:

\begin{enumerate}
\item \emph{Scale-free behavior of degree distribution}:
Let $P(k)$ be the probability that any vertex is connected to $k$ other
vertices (i.e. the vertex is of degree $k$). In many real-world networks,
$P(k)\sim k^{-\gamma}$ with $2\lesssim\gamma\lesssim 3$ \cite{powerlawk}.
 
\item \emph{Clustering property}: 
Clustering of a vertex $c_i$ gives the probability that two nearest
neighbors are connected with each other. And the average clustering coefficient 
$C=\sum_i c_i/N$, where $N$ is the number of vertices in the network, measures the 
global density of interconnected vertex triplets in the network.    
Real-world networks in general exhibit higher clustering coefficient than those
in random networks \cite{smallworld}.

\item \emph{Small-world property}:
A network is said to have small-world property if 
the average shortest path length between two vertices scales
at most logarithmically with $N$ for fixed mean degree \cite{smallworld,smallworld2}.

\item \emph{Correlations of degree}: 
An assortative (A disassortative) network tends to connect vertices with 
similar (dissimilar) degrees. Previous studies found that social networks tend to be assortative 
while technological and biological networks are generally disassortative \cite{cor}.
\end{enumerate}

Various network models have been proposed to simulate or explain the
properties found in complex networks in the real world. For instance, the
Erd\"{o}s-R\'{e}nyi random graph model \cite{random1,random2} generates
static networks that exhibit small-world property.
Nonetheless, the degree distribution of the network generated is
Poissonian rather than a power-law.
Later on, Barab\'{a}si and Albert (BA) extended the Erd\"{o}s-R\'{e}nyi random graph model
by evolving the network using a linear preferential attachment 
mechanism \cite{powerlawk}.
The ideas behind the BA model are that most real-world networks tend to growth 
with time, and a new vertex tends to connect to pre-existing high-degree vertices.  
Sen further incorporated non-linear growth property of the number of links
to the BA model by demanding
the number of links of the vertex added at time $t$ to be
$t^\theta$ for some positive constant $\theta$ \cite{acc}.  
Both the BA model and the Sen model generate 
networks with scale-free topology; but they failed to reproduce the
degree correlations and
clustering properties of real-world networks.

Structural organization of a network can be characterized by
degree-dependent average nearest-neighbors degree $k_{\text{nn}}(k)$
\cite{internet},
the degree-dependent average clustering coefficient $C(k)$ \cite{c(k)}
and assortativity coefficient $r$ \cite{r}.
Pan \emph{et al.}
proposed the generalized local-world
models \cite{xli} in which a subset of vertices is randomly chosen every time
step. The newly added vertex can only make connection to vertices in this
subset. Moreover, Liu \emph{et al.} proposed the self-learning mutual
selection model \cite{jliu} in which every vertex has some probability to self
evolve in each time step. These two models successfully
reproduce the $k_{\text{nn}}(k)$, $C(k)$ and $r$ observed in
real-world networks. 
However, these quantities only 
depend on the topological structure of a network. The relative importance between different
links is not taken into consideration. 
To provide a better description of a network, the intensity of 
interaction among vertices should be taken into account. One may define the
weight of a link $w_{ij}$ as the intensity of interaction 
between vertices $i$ and $j$, 
and the vertex strength $s_i$ as the sum of weight of the links connected to $i$: 
\begin{equation}
 s_i = \sum_{j\in\Gamma (i)} w_{ij} ~,
\end{equation}
where $\Gamma (i)$ denotes the nearest neighbors of $i$.
For instance, the weight $w_{ij}$ in WAN refers to the 
number of seats available on the direct flight connections between the airports
$i$ and $j$. And $s_i$ represents the total traffic going
through the airport $i$.
To probe the weighted networks' architecture, 
a set of weighted quantities are introduced, such as the weighted
degree-dependent average clustering coefficient $C^w (k)$ and 
the weighted degree-dependent average nearest-neighbor degree
$k^w_{\text{nn}}(k)$ \cite{powerlaws}. 
However, no one has proposed a quantity that can directly measure the correlations 
of degree with inclusion of weight so far. 
Here we introduce the weighted assortativity coefficient $r^w$ to measure
the tendency of having a high-weighted link between two vertices of similar
degrees.
In addition, we propose a model that incorporates the 
essential features of strength preferential attachment,
nonlinear growth of number of links, and weight evolution of existing links.
We call this the Weight Evolution Model.
The rule of weight evolution are based on the notion that ``the rich always
gets richer''. In other words, high-weighted link has higher
probability to evolve. Besides, inspired by the work
of Dorogovtsev and Mendes \cite{dm}, existing links in the graph can be
removed. By altering the speed of weight evolution, our model can generate
both assortative and disassortative networks with scale-free behavior
and non-trivial clustering properties. Most importantly, 
both topologically assortative and disassortative networks generated
by our model are in fact weighted assortative.
 
In Section~\ref{model}, we define the Weight Evolution Model. 
We report our analytical calculations and numerical results on the
probability distributions of strength of vertex and 
weight of link in Section~\ref{str}.
Then in Section~\ref{clustering} we define the weighted 
assortativity coefficient and 
compare the organizational structure found in real-world networks 
with our numerical simulations. 
Finally, we give a brief summary in Section~\ref{conclusion}.

\section{Weight Evolution Model} \label{model}
The model starts with a small number $N_0$ of fully connected undirected
vertices. The weight of all the $N_0 (N_0 - 1)/2$ initial links are set to
$w_0$. In each time step, the network evolves under the following rules:

\begin{enumerate}
\item \emph{Topological growth}: A new vertex is added to the network. We
assume that the number of links of the new vertex is an increasing function of
the network size. Specifically, a vertex with $\left[ p
t^\theta \right]$ links are added to the network at time $t$ 
for some fixed constants $p>0$ and $0<\theta<1$.
(Here $[x]$ denotes the value of $x$ rounded to the nearest integer.)
These $\left[ p t^\theta \right]$ links are all of weight
$w_0$ and are randomly connected to the existing vertices according to
the strength preferential probability $\Lambda _i$, 
which is defined as \cite{weighted1,weighted2}
\begin{equation}
 \Lambda_{i} = \frac{s_i}{\sum_j s_j} ~. \label{E:Pro}
\end{equation}	

\item \emph{Evolution of weight}: The link between vertices $i$ and $j$ is chosen to
evolve with probability $Q_{i,j}$ proportional to its weight, namely 
\begin{equation}
 Q_{i,j}=\frac{w_{ij}}{\sum_{a<b} w_{ab}} ~. \label{E:Q}
\end{equation}
The weight will update according to the equation 
\begin{equation}
 w_{ij} \longrightarrow w_{ij} + \text{sgn} (q) ~, \label{E:w_ij_evolution}
\end{equation}
 where
\begin{equation}
 \text{sgn} (q) = \left\{ \begin{array}{cl}
  1 & \text{if~} q > 0 ~, \\
  -1 & \text{if~} q < 0 ~, \\
  0 & \text{if~} q = 0 ~.
 \end{array} \right. \label{E:sgn_Def}
\end{equation}
At time step $t$, this process will repeat $\left[ |q| t^\theta
\right]$ times and a link is removed if the weight equals to $0$. (In other
words, the sign of $q$ determines the addition or subtraction of weight of a
link; and the magnitude of $q$ specifies the rate of the addition/subtraction
per time step.)
\end{enumerate}
Note that $p$, $q$ and $\theta$ are the tunable parameters in this model.

Clearly, our model generalizes the BA model \cite{powerlawk} and the
Sen model \cite{acc}. In particular, our model is reduced to the BA model
by setting $\theta=0$, $q=0$, $w_0=1$, and reduced to
the Sen model by setting $0\le\theta \le 1$, $p=1$, $q=0$ and $w_0=1$.

We now justify the validity of our rules. Note that in most real-world networks,
new vertices are introduced one at a time; and they 
tend to connect to the pre-existing high strength vertices. 
In the example of WAN, new airport tends to
establish flights to existing airports with heavy traffic.  
Our topological growth process is used to capture this kind of network growth.
Moreover, our weight evolution mechanism models the trend that 
the link with higher weight have a higher probability to evolve. For instance, 
in SCN \cite{powerlaws}, a vertex represent a scientist
and a link is present if two scientists have co-authored at least one paper. 
The weight between two scientists is higher if they have higher number of 
co-authored papers.
Naturally, if the weight between two scientists is high, the probability for
them to collaborate again is also high.

\section{Strength and weight} \label{str}
In our model, the processes of topological growth and weight evolution are Markovian. 
Thus, the evolution of our network can be calculated accurately by mean field approximation.
Appendix~\ref{ana} reports the mean field calculation of the probability 
distribution of vertex strength $P(s)$ and probability distribution of weight of link $P(w)$.
We found that $P(s)\sim s^{-\gamma_s}$ and
$P(w)\sim w^{-\gamma_w}$, where

\begin{equation}
 \gamma_s= 1-\left[\frac{2(w_0p+q)}{(\theta-1)w_0p-2q}\right] ~,
 \label{E:gammas}
 \end{equation}	
\begin{equation}
 \gamma_w= 1+ \left[ \frac{w_0p+q}{(\theta+1)q} \right] ~.
 \label{E:gammaw} 
\end{equation}	  

To check the validity of our analytical solution, we  
performed numerical simulations on our network model for different values
of $q$ and $\theta$. Since $p$, $q$ and $w_0$ govern the relative
speed between new link addition
and weight evolution, one can fix two of these three parameters without 
losing any generality. 
In this paper, we fix $p=1$ and $w_0=10$ unless otherwise stated.
Besides, we set the number of initial vertices $N_0$ to $5$; and
we have checked that the
networks generated using different values of $p$ and $N_0$
exhibit similar behaviors as long as $N_0 \ll N$.
Using this set of parameters, we found that statistical data collected
from the network equilibrate after $t\gtrsim 5\,000$. 
Hence, we collect all our statistical data at $t = 10\,000$
and all the data reported here have been averaged over 50 independent runs.
Our model successfully 
reproduces the scale-free behavior of the probability distributions of
strength and weight with a tunable exponent that depends on the 
microscopic mechanism ruling the weight evolution.
We found in Appendix~\ref{num} that the numerical simulation results agree
with the mean field approximation.

\section{Clustering and correlations} \label{clustering}
In this section, we first discuss the quantities used to characterize the 
clustering and correlations behaviors of a network. Then we introduce the weighted 
assortativity coefficient $r^w$. Finally, we report the numerical simulation results 
of our model. 

\subsection{Quantities Characterizing A Weighted Network}
We overload the notation by labeling the vertex added to the network in time
step $i$ also by $i$.
The clustering of $i$ is defined as 
\begin{equation}
c_i= \frac{1}{k_i(k_i-1)}\sum_{j,h}a_{ij}a_{ih}a_{jh}  ~, 
\end{equation} 
where $k_i$ is the degree of $i$ and
\begin{equation}
 a_{ij} = \left\{ \begin{array}{cl}
  1 & \text{if a link exist between $i$ and $j$} ~, \\
  0 & \text{if there is no link between $i$ and $j$} ~.
 \end{array} \right.
\end{equation} 

If vertex $i$ has less than two neighbors, $c_i$ is set to $0$.
Clustering $c_i$ measures the local cohesiveness of vertex $i$
while the average clustering coefficient
$C=\sum_i c_i/N$ measures the global density of interconnected triples in the network. 
Organizational structure of network can be further studied via the degree-dependent average 
clustering coefficient $C(k)$, 
\begin{equation}
 C(k)=\frac{1}{NP(k)}\sum_{i:k_i=k}c_i ~, \label{E:C(k)_Def}
\end{equation}
which is the mean clustering for vertices with degree $k$.
 
To have a better understanding of the organizational structure of weighted networks,  
the weighted clustering $c^w_i$ is introduced and defined as \cite{powerlaws}
\begin{equation}
 c^w_i=\frac{1}{s_i(k_i-1)}\sum_{j,h}\frac{(w_{ij}+w_{ih})}{2}a_{ij}a_{ih}a_{jh}  ~, \label{E:wC(k)_Def}
\end{equation} 
If vertex $i$ tends to form triples with other vertices by its'
high-weighted links, then $c^w_i>c_i$ meaning that 
the topological clustering of $i$ underrates the cohesiveness of $i$.
The weighted average clustering
coefficient $C^w$
and weighted degree-dependent average clustering coefficient $C^w(k)$ are
defined as the average of $c^w_i$ over all vertices and over all vertices with degree
$k$, respectively \cite{powerlaws}.  

The degree correlation is another important information of the network.
Recall that the network is said to be assortative if the vertices tend to 
connect to other vertices which have similar (dissimilar) properties.
Newman introduced the assortativity coefficient \cite{r,cor}
\begin{widetext}
\begin{eqnarray}
r= \frac{M^{-1}\sum_\phi (\prod_{i\in F(\phi)} k_i)-[\frac{M^{-1}}{2}\sum_\phi (\sum_{i\in F(\phi)}k_i)]^2}
{\frac{M^{-1}}{2}\sum_\phi (\sum_{i\in F(\phi)}k_i^2)-[\frac{M^{-1}}{2}\sum_\phi (\sum_{i\in F(\phi)}k_i)]^2}
\end{eqnarray}
\end{widetext}
to measure the degree correlation between linked vertices,
where $F(\phi)$ denotes the set of the two vertices connected by the
$\phi$th link  and $M$ is the total number of links in the network.
This measure $r$ is positive (negative) for assortative (disassortative)
networks; and $r = 0$ for a random graph.
Note that $r$ is independent of the weight of each link of a network.

To probe the degree correlation of the network, one may also study the 
average nearest-neighbors degree, which is defined as \cite{internet} 
\begin{equation}
k_{\text{nn},i}=\frac{1}{k_i}\sum_{j\in\Gamma (i)} k_{j} .
\end{equation}
The degree-dependent average nearest-neighbors degree $k_{\text{nn}}(k)$ 
is the mean of $k_{\text{nn},i}$ restricted to the class of degree $k$ vertices.
In an assortative (disassortative) network, 
vertices with high degree tend to connect to other vertices with high (low) degree,
thus $k_{\text{nn}}(k)$ would be an increasing (decreasing) function of $k$.
In real-world weighted networks, high degree vertices could connect to 
small degree vertices with low weight, 
while connect to high degree vertices with high weight. For instance, in WAN, 
the high degree airport $i$ could have a lot of direct flight to another
high degree airport $j$, while have less number of flight to a low
degree airport $h$. 
In this case, $k_{\text{nn}}(k)$ will underestimate the tendency for
having heavy traffic between two similar degree airports.
To handle this problem, Barrat \emph{et al.} proposed
the weighted average nearest-neighbor degree \cite{powerlaws}
\begin{equation}
k^w_{\text{nn},i}=\frac{1}{s_i}\sum_{j\in\Gamma (i)} w_{ij}k_{j} .
\end{equation}
If the weighted degree-dependent nearest-neighbors degree $k^w_{\text{nn}}(k)$
is an increasing function of $k$, similar degree vertices tend to link
together. Besides, the weights of these links tend to be high. Consequently,
the network is weighted assortative. 

\begin{figure}[t]
 \begin{center}
  \includegraphics*[scale=0.35]{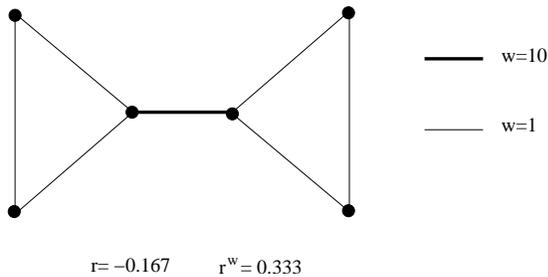}
 \end{center}
  \caption{A network whose $r < 0$ but $r^w > 0$.}
 \label{F:Ex_graph}
\end{figure}

Although one can figure out that the network is assortative or disassortative
if $k^w_{\text{nn}}(k)$ is increasing or decreasing with $k$, a quantity
directly describing the weighted assortativity is needed. 
Thus we introduce the weighted assortativity coefficient
\begin{widetext}
\begin{eqnarray}
r^w= \frac{H^{-1}\sum_\phi
(\varpi_\phi\prod_{i\in F(\phi)} k_i)-[\frac{H^{-1}}{2}\sum_\phi
(\varpi_\phi\sum_{i\in F(\phi)}k_i)]^2}
{\frac{H^{-1}}{2}\sum_\phi (\varpi_\phi\sum_{i\in F(\phi)}k_i^2)-
[\frac{H^{-1}}{2}\sum_\phi (\varpi_\phi\sum_{i\in F(\phi)}k_i)]^2} ~,
\end{eqnarray}
\end{widetext}
where $\varpi_i$ is the weight of the $\phi$th link, 
$F(\phi)$ is the set of the two vertices connected by the $\phi$th link  
and $H$ is the total weight of all links in the network.
Just like $r$, $r^w$ lies between $-1$ and $1$. Moreover, $r^w$ 
is positive for weighted assortative networks, while
negative for weighted disassortative networks.
If $r^w > r$, a high-weighted link tend to connect two similar degree vertices
together.
In fact, $r^w$ reduces to $r$ if all weights in the network are equal.
Furthermore, $r^w = r = 0$ for a random graph.
Fig.~\ref{F:Ex_graph} illustrates that the
values of $r$ and $r^w$ of a network can differ greatly. The network
in Fig.~\ref{F:Ex_graph} consists of two complete graphs of three vertices
that are joined together by a highly weighted link.
Since the five out of seven of the links in this network connect two vertices
of different degree together, the assortativity coefficient $r$ of this
network is negative. (In fact $r = -1/6$.)
On the other hand, the weighted assortativity coefficient $r^w$ of this network 
is positive. This can be understood as follows.
The link between the two degree three vertices in this network carries the most
weight. So after coarse-graining, the network
is similar to the one making up of a single link connecting two degree one
vertices together plus another four isolated vertices. 
And clearly, the coarse-grained network is assortative.
Therefore, when the weight of link are taken into account, the network in
Fig.~\ref{F:Ex_graph} ought to be assortative rather than disassortative.
(In fact $r^w = 2/3$.)
In this respect, the weighted assortativity coefficient $r^w$ better determines
the assortativity of networks with weighted links than the assortativity
coefficient $r$.

\begin{figure}[t]
 \begin{center}
  \includegraphics*[scale=1]{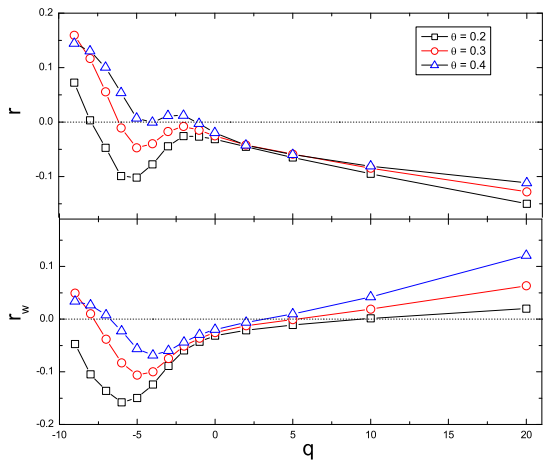}
 \end{center}
  \caption{Assortativity coefficient $r$ and weighted assortativity
   coefficient $r^w$ versus $q$ for various $\theta$.
The statistics taken right at time $t=10\,000$.
And we have averaged over $50$ independent runs in all the data
reported here. All error bars are $\lesssim 10^{-2}$.}
 \label{F:r}
\end{figure}

\begin{figure}[t]
 \begin{center}
  \includegraphics*[scale=1]{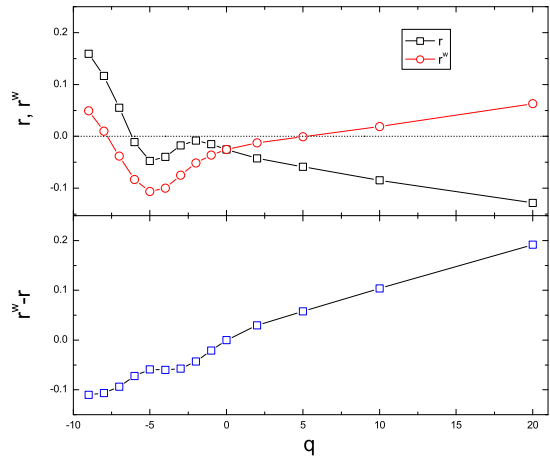}
 \end{center}
  \caption{The comparison between $r$ and $r^w$ for $\theta=0.3$.}
 \label{F:rw-r}
\end{figure}

\subsection{Simulation Results}
As shown in Fig.~\ref{F:r}, our model can generate assortative networks ($r>0$),
such as those found in social networks, when $q\ll 0$.
Our model can also generate disassortative networks ($r<0$), such as those found
in technological networks, when $q\gg 0$.
Surprisingly, $r^w > 0$ for both assortative and disassortative networks provided
that $|q| \gtrsim 7$. This observation indicates
that the topological assortativity coefficient underrates the contributions of
high-weighted links between similar degree vertices.
Our finding shows that studying the assortativity of a network by
considering network topology alone \cite{r,cor} does not give a complete
picture of the organizational structure of the network.

The difference between $r^w$ and $r$ (Fig.~\ref{F:rw-r}) can be understood qualitatively
by considering the evolution mechanisms of our model. For $q\gg 0$, a newly
added vertex is the one with low degree and it tends to connect to
existing high-strength and high-degree vertices.
This process leads to
topologically disassortative behavior. Meanwhile, a link introduced at an early
time, which connects two old vertices having similar degrees together, tends to be
high-weighted as $q>0$.
This leads to the observed weighted assortative behavior.    
On the other hand, assortative networks emerge at $q\ll 0$. It is because the 
strength of old vertices are low. Hence, there is a high chance for two young
vertices having similar degrees to connect.
As $\theta$ increases, the strength of old vertices decrease
while the new vertices enter the network with a higher strength.
Thus, for $q\ll 0$, the value of $|q|$ required for the emergence of
assortative behavior decreases with $\theta$. This is precisely what we have
observed in Fig.~\ref{F:r}.

\begin{figure}[t]
 \begin{center}
  \includegraphics*[scale=1]{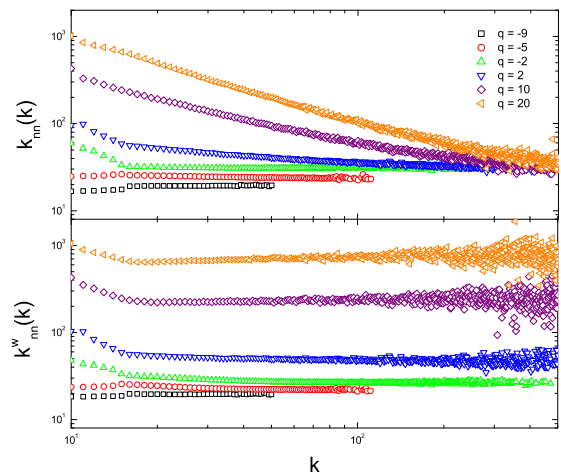}
 \end{center}
  \caption{$k_{\text{nn}}(k)$ and $k^w_{\text{nn}}(k)$ for various $q$ for
  $\theta=0.3$.}
 \label{F:k}
\end{figure}

\begin{figure}[t]
 \begin{center}
  \includegraphics*[scale=1]{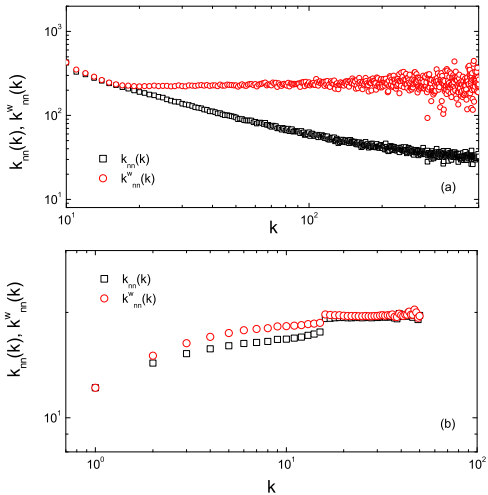}
 \end{center}
  \caption{Comparison of $k_{\text{nn}}(k)$ and $k^w_{\text{nn}}(k)$ for (a)
   $q=10$ and (b) $q=-9$ for $\theta=0.3$.}
 \label{F:k2}
\end{figure}

\begin{figure}[t]
 \begin{center}
  \includegraphics*[scale=1]{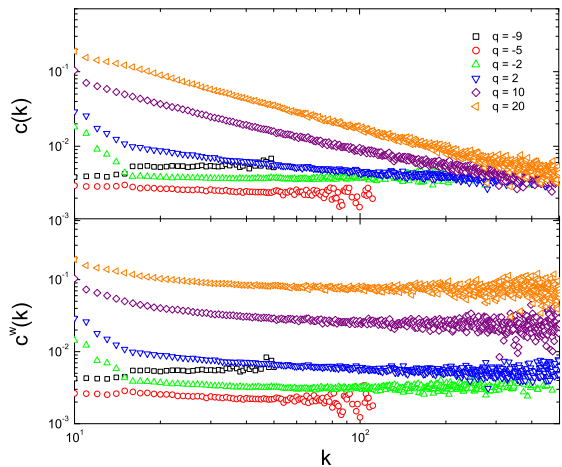}
 \end{center}
  \caption{$C(k)$ and $C^w(k)$ for various $q$ for $\theta=0.3$.}
 \label{F:C(k)}
\end{figure}

As shown in Fig.~\ref{F:k}, $k_{\text{nn}}(k)$ exhibits decreasing power-law behavior for
$q\gg 0$, indicating that the networks are disassortative. For $q \ll 0$, 
$k_{\text{nn}}(k)$ is a gently increasing function of $k$. It is worth noting
that in 
some empirical studies of real-world networks \cite{powerlaws}, both $k_{\text{nn}}(k)$ and $k^w_{\text{nn}}(k)$
show assortative behavior with $k^w_{\text{nn}}(k)>k_{\text{nn}}(k)$.
As shown in Fig.~\ref{F:k2},
our model successfully reproduces these important features.
Remarkably, $r^w<r$ while $k^w_{\text{nn}}(k)>k_{\text{nn}}(k)$ for $q\ll 0$. 
This points out that it is difficult to compare the weighted assortativity and topological 
assortativity by studying the difference between $k^w_{\text{nn}}(k)$ and
$k_{\text{nn}}(k)$.

Two types of $C(k)$ are found in real-world networks \cite{c(k),class}. 
In the first type, $C(k)$ does not exhibit strong dependency on $k$. 
While in the second type, $C(k)$ shows a decreasing power law spectrum. 
Fig.~\ref{F:C(k)} clearly shows that our model can reproduce both kinds of
features.
In particular, $C(k)$ is flat for $q\ll 0$ and it becomes a decreasing power
law for $q\gg 0$. 
Moreover, $C^w(k)$ is larger than $C(k)$, especially at large $k$, which is a feature
found in real-world networks \cite{powerlaws}. This result indicates that 
the topological clustering
coefficient $C(k)$ underestimate the cohesiveness of the network. It is
because $C(k)$ cannot tell us whether
large degree vertices tend to form interconnected
triples with its' high-weighted links. 

\section{Conclusions} \label{conclusion}
In summary, we have introduced the Weight Evolution Model which
couples dynamical evolution of weight with topological network growth.
We have also introduced the weighted assortativity coefficient $r^w$ to
characterize the degree correlation of a weighted network.
Our model reproduced many features found in real-world networks, such as
scale-free behavior, assortative and disassortative behavior, and
two types of clustering structures. 
Most importantly, both topologically disassortative and assortative networks are 
regarded as weighted assortative if the weight of links is taken into account,
indicating that the topological assortativity coefficient $r$ 
underrates the contributions of high-weighted links between similar
degree vertices.
This result may give us some idea for why
social networks are found to be assortative, 
while technological networks and biological networks show an
opposite behavior in previous studies \cite{r,cor}.
It is instructive to investigate if all real-world networks can be regarded as
weighted assortative.

\acknowledgments
 We would like to thank the Computer Center of HKU for their helpful support
 in providing the use of the HPCPOWER~System for the simulation reported in
 this paper. Useful discussions with F.~K. Chow, K.~H. Ho and V.~H. Chan are
 gratefully acknowledged.

\appendix
\section{Analytical calculations of strength and weight}\label{ana}
Recall that initially the network is a complete graph of $N_0$ vertices.
And in each time step a new vertex is added.
There are two ways to alter the strength of an existing vertex $i$:
(a)~$i$ is connected to the newly added vertex $j$ with probability 
$\Lambda_i$ given by Eq.~(\ref{E:Pro}), or (b)~the 
weight of the link between $i$ and one of its neighboring vertices evolves
according to Eqs.~(\ref{E:Q})--(\ref{E:sgn_Def}).

By mean field approximation and by treating all discrete 
variables as continuous, the strength $s_i$ of the vertex added at time $i$ satisfies
\begin{eqnarray}
\frac{ds_i(t)}{dt}&=& w_0 p
t^\theta \frac{ s_i(t)}{\sum_\ell s_\ell (t)} + q t^\theta
\frac{\sum_{j\in\Gamma(i)} w_{ij}(t)}{\sum_{a<b} w_{ab}(t)} \nonumber \\
&=& w_0 p t^\theta \frac{ s_i(t)}{\sum_\ell s_\ell (t)} + 2 q t^\theta
\frac{s_i(t)}{\sum_\ell s_\ell (t)} ~, \label{E:ds_i}
\end{eqnarray}	
for $t > i$.
Note that the first and second terms represent the strength change due to
topological growth and evolution of weight, respectively.

Since the total strength of the network increases $2(w_0 p + q)t^\theta$
approximately in each time step, 
\begin{equation}
\sum_l s_l(t) \approx \int_0^t 2(w_0p+q)t^\theta dt = \frac{2(w_0p+q)}{\theta
+1}\,t^{\theta +1} ~. \label{E:sum_s_approx}
\end{equation}	
Putting Eq.~(\ref{E:sum_s_approx}) into Eq.~(\ref{E:ds_i}) and using the initial
condition $s_i (t=i)= w_0 [p i^{\theta}]$, we conclude that
\begin{eqnarray}
 s_i(t)&\approx& w_0 p i^{\theta}
  \left(\frac{t}{i}\right)^{(\theta+1)(w_0p+2q)/2(w_0p+q)} \nonumber \\
 &=& w_0 p t^{\theta} \left(\frac{t}{i}\right)^{[2q-(\theta-1)w_0p]/2(w_0p+q)} ~.
 \label{E:s(t)}
\end{eqnarray}	

Since one vertex is added to the network in each time step,
the network size $N{\approx}t$.
The probability distribution of vertex strength can be computed by
\begin{equation}
 P(s)\approx \frac{1}{t}\int_{0}^{t} \delta(s-s_j(t))dj ~,
 \label{E:p(s)}
\end{equation}	
where $\delta(x)$ is the Dirac delta function.  
Combined with Eq.~(\ref{E:s(t)}), we conclude that in the infinite network size
limit, the mean field approximation give $P(s)\sim s^{-\gamma_s}$ for 
$\textrm{min}(s_i(t))\le s\le \textrm{max}(s_i(t))$, where
\begin{equation}
 \gamma_s= 1-\left[\frac{2(w_0p+q)}{(\theta-1)w_0p-2q}\right] ~.
 \label{E:gammas_alt}
\end{equation}	 

\begin{figure}[ht]
 \begin{center}
  \includegraphics*[scale=1]{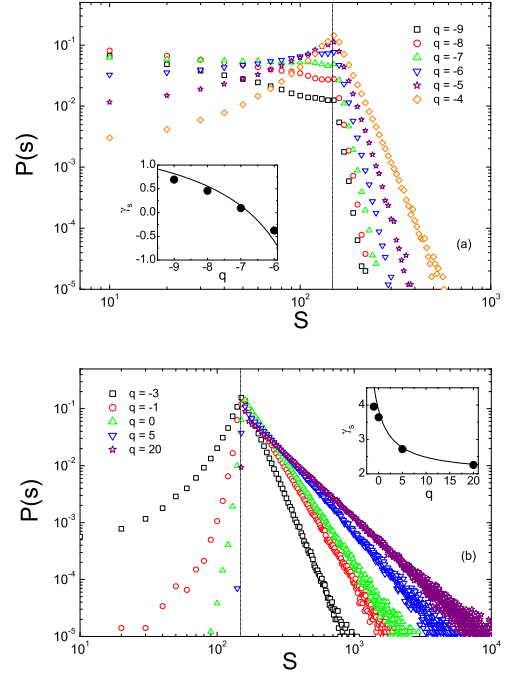}
 \end{center}
 \caption{Probability distribution of strength $P(s)$ for $\theta=0.3$.
A power-law function $P(s) \sim s^{-\gamma}$ is obtained in the range of
(a) $s<s_n$ if $q<q_c=-3.5$ 
and (b) $s>s_n$ if $q>q_c$.
The dashed line corresponds to $s=s_n=148$.
The insert compares the 
values of $\gamma$ obtained from data fitting
(filled circles) with the analytical solution $\gamma_s$ given
by Eq.~(\ref{E:gammas_alt}) (solid line).}
 \label{F:P(s)}
\end{figure}

The behavior of $P(s)$ can be classified according to the value of $q$: 
\begin{enumerate}
\item If $q>q_c\equiv(\theta-1)w_0p / 2$, Eq.~(\ref{E:s(t)}) tells us that
the degree of existing vertices are generally larger than the degree of
the newly added vertex, i.e. $s_i(t)\ge s_n \equiv w_0pt^{\theta}$ for all $i$.
Thus,
\begin{equation}
 P(s)  \sim \left\{ \begin{array}{lll}
   &0  &\text{for~} s < s_n ~, \\
   & s^{-\gamma_s} & \text{for~} s \ge s_n~.
 \end{array} \right.
\end{equation}
In particular, the network evolves purely by topological
growth for $q=0$. And in this case, we
find that $\gamma_s=(\theta-3)/(\theta-1)$ which is independent of $p$.
Moreover, $\gamma_s\rightarrow 2$ as $q\rightarrow\infty$.
 
\item If $q=q_c$, $s_i(t)=s_n$ 
which is the same for all $i$. In this case, 
$P(s)$ is a delta function. In contrast, $P(s)$ is a delta function only
when $\theta = 1$ (\emph{i.e.}, when the network is fully connected) in the
Sen model \cite{acc}.

\item If $q<q_c$, 
$s_i(t) \le s_n$ by Eq.~(\ref{E:s(t)}). In other words,
the newly added vertex have the highest strength. Using a similar argument as
in the case of $q>q_c$, we found that the strength distribution
$P(s)$ follows
\begin{equation}
 P(s)  \sim \left\{ \begin{array}{lll}
 & s^{-\gamma_s} & \text{for~} s < s_n~,\\
 & 0 &\text{for~} s \ge s_n ~.
 \end{array} \right.
\end{equation}
\end{enumerate}

The evolution of weight can again be computed using mean field approximation.
Therefore, we have
\begin{equation}
\frac{dw_{ij}(t)}{dt} \approx q t^\theta \frac{w_{ij}(t)}{\sum_{a<b}
w_{ab}(t)} \approx \frac{(\theta+1) q w_{ij}(t)}{(w_0 p + q) t} ~, \label{E:dw_ij}
\end{equation}	 
for $t > \max (i,j)$. Clearly, $w_{ij}(t)$ satisfies the initial
condition $w_{ij}(t=\text{max}(i,j))=w_0$. Hence, we find
\begin{equation}
 w_{ij}(t) \approx w_0\left[ \frac{t}{\textrm{max}(i,j)}
 \right]^{(\theta+1)q/(w_0 p +q)} ~.
 \label{E:w(t)}
\end{equation}	
 
\begin{figure}[t]
 \begin{center}
  \includegraphics*[scale=1]{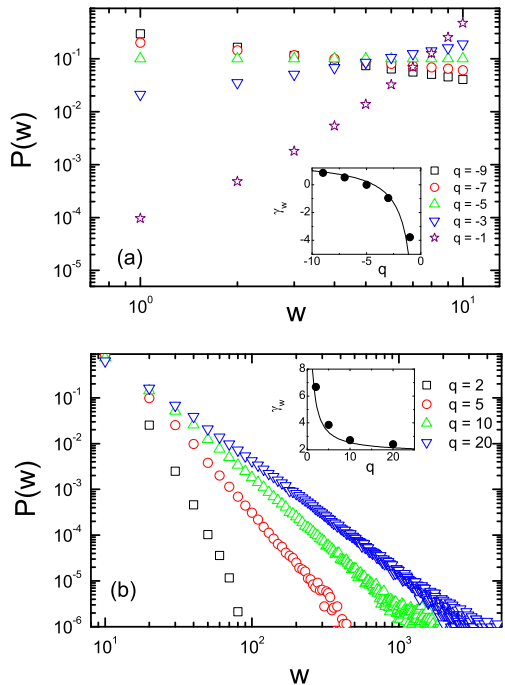}
 \end{center}
  \caption{Probability distribution of weight $P(w)\sim w^{-\gamma _w}$ with
(a) $q<0$ and (b) $q>0$.
The insert compares the values of $\gamma_w$ obtained from data fitting
(filled circles) and the mean field value given by Eq.~(\ref{E:gammaw_alt})
(solid line) for $\theta=0.3$.}
 \label{F:P(w)}
\end{figure}

In addition, the weight distribution $P(w)$, which gives the probability that
a link with weight $w$, obeys $P(w)\sim w^{-\gamma_w}$ for $\textrm{min}(w_{ij})\le w\le \textrm{max}(w_{ij}) $, where
\begin{equation}
 \gamma_w= 1+ \left[ \frac{w_0p+q}{(\theta+1)q} \right] ~.
 \label{E:gammaw_alt}
\end{equation}	  

\section{Numerical results of probability distribution} \label{num}
Fig.~\ref{F:P(s)} reports $P(s)$  
obtained from numerical simulations for different values of $q$. It also
compares the value of exponent obtained from fitting the numerically
simulated data with the value predicted by
analytical calculation. $P(s)$ follows a power-law function for $q\ll q_c$ or 
$q\gg q_c$ with exponent agree well with the value given by Eq.~(\ref{E:gammas_alt}).
Note that $P(s)$ is non-zero for $s>s_n$ when $q<q_c$ (Fig.~\ref{F:P(s)}(a)) and
for $s<s_n$ when $q>q_c$ (Fig.~\ref{F:P(s)}(b)).
This discrepancy is due to the fact that a small proportion of vertices
evolves differently from the prediction of mean-field
approximation. However, this value diminishes to 
zero rapidly and thus our analytical calculation remains valid. 
The weight probability distribution $P(w)$ for various $q$ is displayed in Fig.~\ref{F:P(w)} 
along with the comparison between the fitted values of the exponent with 
the analytical predictions. For $q<0$, $w_{ij}\le w_0$ for all links, so the power-law 
function is obtained for $w\le w_0$ (Fig.~\ref{F:P(w)}(a)). For $q>0$, the power-law
function is found to be in the range of $w\ge w_0$ (Fig.~\ref{F:P(w)}(b)).


\begin{thebibliography}{99}
\bibitem{review1} S.~N.~Dorogovtsev and J.~F.~F.~Mendes, Adv. Phys.
 \textbf{51}, 1079 (2002).
\bibitem{review2} R.~Albert and A.~L.~Barab\'{a}si, Rev. Mod. Phys.
 \textbf{74}, 47 (2002).
\bibitem{review3} M.~E.~J.~Newman, SIAM Rev. \textbf{45}, 167 (2003).
\bibitem{internet} R.~Pastor-Satorras, A.~V\'azquez, and A.~Vespignani,
Phys. Rev. Lett. \textbf{87}, 258701 (2001).
\bibitem{internet2} A.~V\'azquez, R.~Pastor-Satorras, and A.~Vespignani,
Phys. Rev. E \textbf{65}, 066130 (2002).
\bibitem{internet3} R.~Pastor-Satorras and A.~Vespignani,
\textit{Evolution and Structure of the Internet:
A Statistical Physics Approach} (Cambridge University Press, Cambridge,
U.K., 2004).
\bibitem{www} R.~Albert, H.~Jeong, and A.-L.~Barab\'{a}si, Nature
 \textbf{401}, 130 (1999).
\bibitem{scn} M.~E.~J.~Newman, Proc. Natl. Acad. Sci. U.S.A. \textbf{98},
 404 (2001). 
\bibitem{scn2} A.-L.~Barab\'{a}si, H.~Jeong, Z.~N\'{e}da, E.~Ravasz,
 A.~Schubert, and T.~Vicsek, Physica A \textbf{311}, 590
 (2002).
\bibitem{bio} H.~Jeong, S.~P.~Mason, A.-L.~Barab\'{a}si and Z.~N.~Oltvai, Nature \textbf{411}, 41 (2001).
\bibitem{bio2} E.~Ravasz, A.~L.~Somera, D.~A.~Mongru, Z.~N.~Oltvai,
and A.-L. Barab\'{a}si, Science \textbf{297}, 1551 (2002).
\bibitem{airport} R.~Guimer\`{a}, S.~Mossa, A.~Turtschi, and L.~A.~N.~Amaral,
Proc. Natl. Acad. Sci. U.S.A. \textbf{102}, 7794 (2005).
\bibitem{powerlawk} A.-L.~Barab\'{a}si and R.~Albert,
 Science \textbf{286}, 509 (1999).
 \bibitem{smallworld} D.~J.~Watts and S.~H.~Strogatz, Nature
 \textbf{393}, 440 (1998).
 \bibitem{smallworld2} M.~E.~J.~Newman, Phys. Rev. E \textbf{64}, 016132 (2001).
\bibitem{cor} M.~E.~J.~Newman, Phys. Rev. E \textbf{67}, 026126 (2003).
\bibitem{random1} P.~Erd\"os and A.~R\'{e}nyi. Publ. Math. (Debrecen) \textbf{6}, 290 (1959).
\bibitem{random2} B.~Bollob\'as, \textit{Random Graphs} (Academic Press,
 London, 1985), chap~2.
\bibitem{acc} P.~Sen, Phys. Rev. E. \textbf{69}, 046107 (2004).
 \bibitem{c(k)} E.~Ravasz and A.-L.~Barab\'{a}si, Phys. Rev. E
 \textbf{67}, 026112 (2003).
\bibitem{r} M.~E.~J.~Newman, Phys. Rev. Lett. \textbf{89}, 208701 (2002).
 \bibitem{xli} Z.~Pan, X.~Li, and X.~Wang, Phys. Rev. E \textbf{73}, 056109
 (2006).
 \bibitem{jliu} J.-G.~Liu, Y.-Z.~Dang, W.-X.~Wang, Z.-T.~Wang, T.~Zhou,
 B.-H.~Wang, Q.~Guo, Z.-G.~Xuan, S.-H.~Jiang and M.-W.~Zhao,
 {\tt arXiv:physics/0512270} (2005).
\bibitem{powerlaws} A.~Barrat, M.~Barth\'elemy, R.~Pastor-Satorras, and
 A.~Vespignani, Proc. Natl. Acad. Sci. U.S.A. \textbf{101}, 3747 (2004).
 \bibitem{dm} S.~N.~Dorogovtsev and J.~F.~F.~Mendes,
 Europhys. Lett. \textbf{52}, 33 (2000).
 \bibitem{weighted1} A.~Barrat, M.~Barth\'elemy,
 and A.~Vespignani, Phys. Rev. Lett. \textbf{92}, 228701 (2004).
 \bibitem{weighted2} A.~Barrat, M.~Barth\'elemy,
 and A.~Vespignani, Phys. Rev. E. \textbf{70}, 066149 (2004).
 \bibitem{class} A.~V\'azquez, Phys. Rev. E. \textbf{67}, 056104 (2003).
\end{thebibliography}
\end{document}